\documentclass[onecolumn,preprintnumbers,amssymb,amsmath,superscriptaddress,letterpaper,nofootinbib]{revtex4}[12pt]
\usepackage{times,graphics}
\usepackage[colorlinks]{hyperref}

\usepackage{graphicx}
\usepackage{dcolumn}
\usepackage{bm}
\usepackage{natbib}
\usepackage{pstricks}
\usepackage{epsfig}
\usepackage{epstopdf}
\usepackage{multirow}

\newcommand{\be}{\begin{equation}}
\newcommand{\ee}{\end{equation}}
\newcommand{\bea}{\begin{eqnarray}}
\newcommand{\eea}{\end{eqnarray}}

\makeatletter
\newcommand*{\shifttext}[2]{%
	\settowidth{\@tempdima}{#2}%
	\makebox[\@tempdima]{\hspace*{#1}#2}%
}
\makeatother

\newcommand{\metalambda}{%
	\mathop{%
		\rlap{$\Lambda$}%
		\mkern2mu
		\shifttext{2.5pt}{$\Lambda$}%
	}%
}

\begin{document}

\title{``Ups and Downs in Dark Energy"\\
phase transition in dark sector as a proposal to lessen $H_0$ tension}
\author{Abdolali Banihashemi}
\email{a\_banihashemi@sbu.ac.ir}
\affiliation{Department of Physics, Shahid Beheshti University, G.C., Evin, Tehran 19839, Iran}

\author{Nima Khosravi}
\email{n-khosravi@sbu.ac.ir}
\affiliation{Department of Physics, Shahid Beheshti University, G.C., Evin, Tehran 19839, Iran}

\author{Amir H. Shirazi}
\email{amir.h.shirazi@gmail.com }
\affiliation{Department of Physics, Shahid Beheshti University, G.C., Evin, Tehran 19839, Iran}

\date{\today}

\begin{abstract}
Based on tensions between the early and late time cosmology, we proposed a double valued cosmological constant which could undergo a phase transition in its history. It is named ``double-$\Lambda$ Cold Dark Matter": $\metalambda$CDM. An occurred phase transition results in (micro-) structures for the dark sector with a proper (local) interaction. In this paper, inspired by the physics of critical phenomena, we study a simplified model such that the cosmological constant has two values before a transition scale factor, $a_t$, and afterwards it becomes single-valued. We consider both the background and perturbation data sets including CMB, BAO distances and R19 data point. $\metalambda$CDM has  its maximum likelihood for $a_t= 0.916^{+0.055}_{-0.0076}$ and  $H_0= 72.8\pm 1.6$. This result shows no inconsistency  between early and late time measurements of Hubble parameter in $\metalambda$CDM model. In comparison to $\Lambda$CDM, our model has better fit to data such that $\Delta\chi^2=-11$ and even if we take care of two additional degrees of freedom we do have better AIC quantity $\Delta$AIC$=-7$. We conclude that a phase transition in the behavior of dark energy can address $H_0$ tension successfully and may be responsible for the other cosmological tensions.

\end{abstract}
\maketitle
\section{Introduction:}

The standard model of cosmology, $\Lambda$CDM,  is very successful in describing the cosmological data from the early universe \cite{Aghanim:2018eyx,Ade:2015xua} as well as the late time observations \cite{Tegmark:2003ud}. Its constituents are cold dark matter (CDM) and the cosmological constant, $\Lambda$. CDM and $\Lambda$ are responsible for matter structure formation and the late time acceleration phase, respectively. However due to mysterious (dark) nature of its main elements, it is a relevant question to ask if dark matter and dark energy are fundamental or not. On the other hand both theoretically and observationally there are few issues which should be answered in the context of $\Lambda$CDM. One of the outstanding (theoretical) question is the cosmological constant fine-tuning problem \cite{CC-problem}. On the other hand, recently, some tensions have been reported between $\Lambda$CDM predictions and the observations. To address these issues there are different approaches which go beyond standard $\Lambda$CDM. We think these tensions can be phrased as follows:  a $\Lambda$CDM which its free parameters are fixed by early universe data (mainly CMB) is not consistent (up to few $\sigma$'s) with a $\Lambda$CDM which is constrained by late time observations (i.e. LSS data). A recent work in this direction claims that dynamical dark energy is favored by $3.5\,\sigma$  over $\Lambda$CDM \cite{Zhao:2017cud}. Their approach is interesting because they look for the dark energy equation of state by reconstructing it directly from the observational data \cite{Zhao:2017cud,Wang:2018fng}.

The most famous tension is $H_0$ tension which is between measurements of Hubble parameter at $z=0$ by CMB \cite{Aghanim:2018eyx} and supernovae \cite{Riess:2016jrr,riess18,Riess:2019cxk,Bernal:2016gxb} where late time direct measurement predicts higher value for $H_0$ in comparison to Planck 2018. This tension was reported in the literature and became worse with the recent measurements \cite{Riess:2019cxk} although it can be a systematic error in the observations. In this paper we focus on this tension. However, there are other (mild) tensions like $f\,\sigma_8$ tension  where again the measurement of matter density between late time observations \cite{Abbott:2017wau} and CMB \cite{Aghanim:2018eyx} are not compatible; or BAO Lyman-$\alpha$ \cite{Bourboux:2017cbm}, void phenomenon \cite{Peebles:2001nv} and missing satellite problem \cite{Klypin:1999uc} where the last two ones are in non-linear regime. On the theory side, there are different strategies to address these tensions but all of them need to go beyond standard model of cosmology. An interesting candidate for this purpose is massive neutrinos but it cannot address both $H_0$ and $f\,\sigma_8$ tensions together \cite{Aghanim:2018eyx}. There are also other ideas trying to solve either $H_0$ or $f\,\sigma_8$ tensions e.g. early dark energy \cite{Poulin:2018cxd} interacting dark energy \cite{DiValentino:2017iww,Yang:2018euj,DiValentino:2019jae,DiValentino:2019ffd}, negative cosmological constant \cite{Visinelli:2019qqu}, late time decaying dark matter \cite{Vattis:2019efj}, neutrino-dark matter interaction \cite{DiValentino:2017oaw,Ghosh:2019tab},  varying Newton constant \cite{Nesseris:2017vor}, viscous bulk cosmology \cite{Mostaghel:2016lcd}, massive graviton \cite{DeFelice:2016ufg} and many more. Recently, another idea, named \"u$\Lambda$CDM, has been studied in the literature to address $H_0$ tension by assuming two different behavior in high and low redshifts \cite{Khosravi:2017hfi} which is very similar to \cite{DiValentino:2017rcr}. This model is based on some theoretical motivations \cite{Khosravi:2016kfb,Khosravi:2017aqq}. In \"u$\Lambda$CDM, cosmological model swtiches, at a transition redshift $z_t$, from the standard $\Lambda$CDM model  to $R=R_0$ model, where $R$ is the Ricci scalar and $R_0$ is a constant. This feature of \"u$\Lambda$CDM model brings us to a new idea to solve the cosmological tensions.

Before discussing this idea let us repeat that it seems all of the cosmological tensions have the same format if we phrase them as: the physics of late time is different from the early universe physics. According to this viewpoint we suggest a new concept/idea in the physics of cosmological models: phase transition in dark sector. In this work we pursue this idea that a phase transition has happened in the dark sector (here we focus on dark energy). The reason for this can be a (microscopical) structure in dark energy like a kind of spin, for example. The idea of phase transition has been studied extensively under a more general topic: critical phenomena.

Critical phenomena are revisited in a variety fields of physics, which local interactions of a many-body system result in a global phase transition. Usually, an ordered phase emerges by lowering the free parameter of the model, e.g. temperature, beyond a critical point. For example, Ising model is classic model of critical phenomena, which describes the phase transition from para-magnet to ferro-magent at Curie temperature. It consists of two-directions magnetic dipoles, i.e. spins, which interacts with each other on a lattice and enforce their neighbors to align with them. In high temperature regime, spins take directions randomly regardless of their neighbors' directions. Close enough to the critical temperature, however, neighbor interactions result in the emergence of aligned islands. Consequently, there is one dominant direction at low temperature regime.

In the next section we propose a model inspired by Ising model for dark energy which (possibly) experiences a phase-transition. Based on the idea of this work we have studied a more general scenario in \cite{Banihashemi:2018has} but without including CMB dataset. In the section \ref{datasets} we constrain our model's free parameters with CMB, BAO and R19 data sets. In section \ref{conclusions} we will conclude and give future perspective on our idea in section \ref{future}. In appendix \ref{at} we will discuss the behavior of 1-d marginalized likelihood for $a_t$ and explain our motivations for limiting the prior for this parameter.

\section{$\metalambda$CDM Model}\label{model}

\begin{figure}
	\centering
	\includegraphics[width=0.7\linewidth]{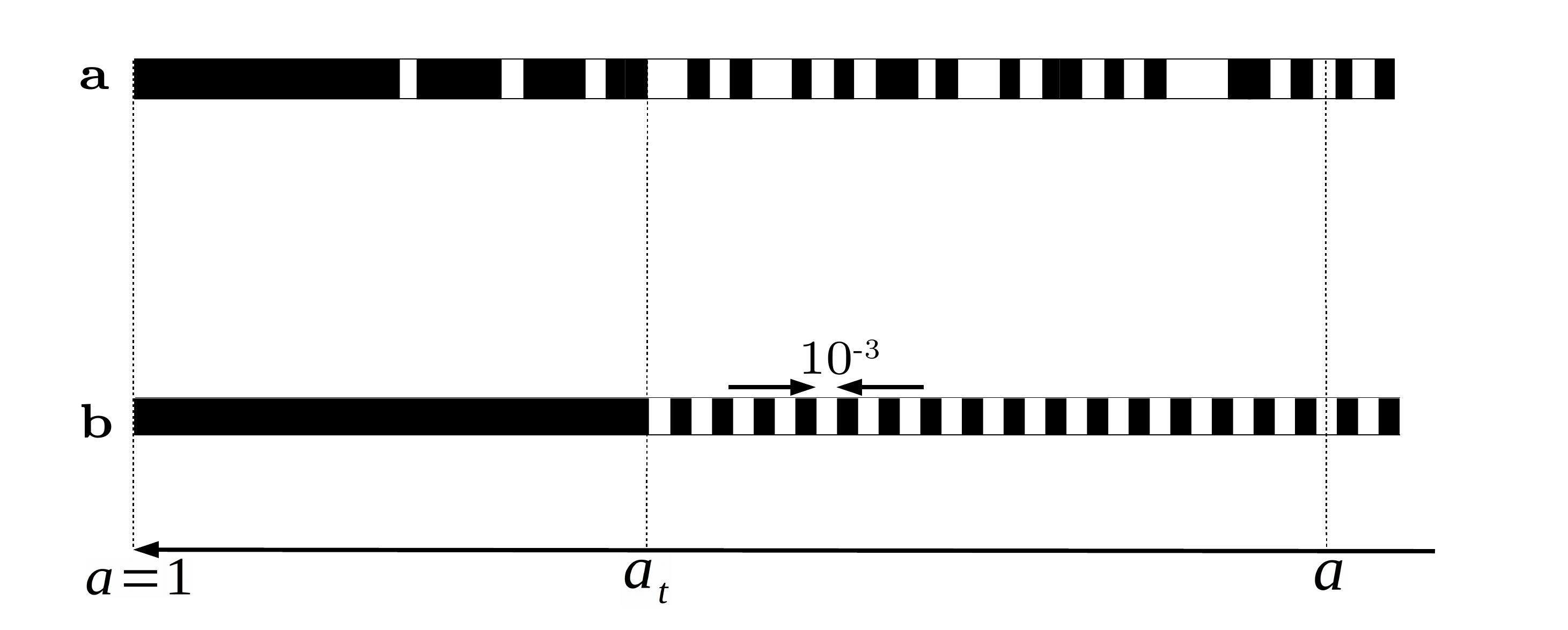}
	\caption{Here we sketch Ising model and our approximation of it in a cartoon. In line \textbf{a} we can see that Ising model in high temperatures sees two states (here we showed them by black and white small boxes) randomly. However when we are close to the critical temperature then one of the states becomes dominant (here the black states). Then if the temperature goes to absolute zero all the states will be black. In our approximation, line \textbf{b}, we assumed before the transition scale factor we have black and white states one by one and after the transition scale factor we switch to black states. Note that in our cosmological scenario increasing scale factor $a$ means decreasing temperature and  a transition scale factor $a_t$ corresponds to the critical temperature.}
	\label{fig:demonst}
\end{figure}

\begin{figure}
	\centering
	\includegraphics[width=0.7\linewidth]{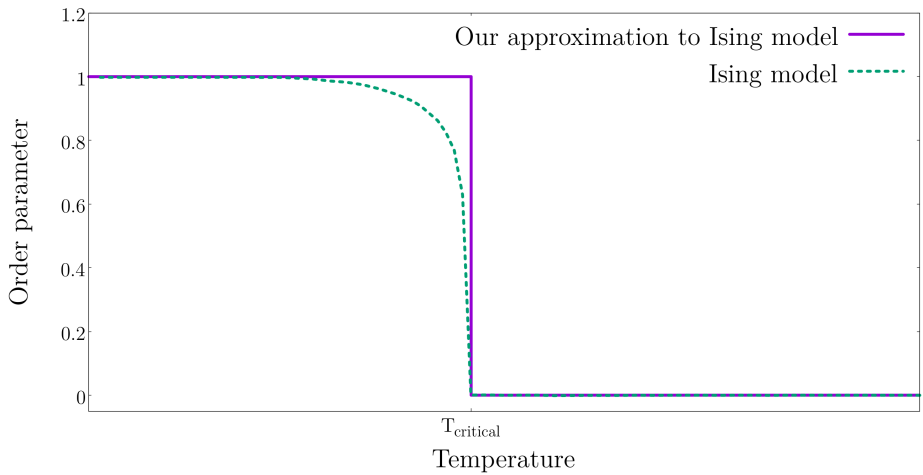}
	\caption{We have plotted the order parameter versus temperature. In a real Ising model the order parameter (in this case the magnetization)  which is zero for above $T_c$ starts to take either a positive or negative value. In our approximation this transition is assumed to be sharp as it is demonstrated in FIG. \ref{fig:demonst}. Physically, it means our system transits from the critical temperature very quickly.}
	\label{fig:order}
\end{figure}
We realize  a phase transition behavior in dark energy sector by an inspiration from Ising model. In the Ising model two-valued spin is at work and a local interaction between these two spins govern the behavior of the system. By reducing the temperature the system can go either to almost spin-up or spin-down state if the temperature becomes less than a critical temperature, $T_c$, and for the absolute zero temperature all the spins will be aligned as we have shown in line \textbf{a} in FIG. \ref{fig:demonst} schematically.  In the cosmology framework we assume the dark energy sector has a structure and to realize it instead of spin we assume a two-valued cosmological constant. We should emphasize that for our purposes a 3-D Ising simulation with enough resolution is practically impossible. So we decided to simplify the model to make it computationally affordable though we keep the interesting properties of it. We assume above the critical temperature spin-up and spin-down states appears one by one but below the critical temperature all the spins are aligned. More precisely it means we assumed the temperature dependence of the order parameter has a sharp behavior instead of a smooth one, FIG. \ref{fig:order}. In this setup we will have a transition scale factor, $a_t$, corresponds to $T_c$. This is a natural choice due to the relation between universe temperature and scale factor $T\propto 1/a$. Note that we assumed the two-valued cosmological constant sees photon thermal bath which is an assumption. The universe sees both values of $\Lambda_1$ and $\Lambda_2$ before $a_t$ but after the phase transition at $a=a_t$ everything switches to just one of these values for $\Lambda$, either $\Lambda_1$ or $\Lambda_2$ $^,$\footnote{Note that having different cosmological constant in different redshift has been considered in \cite{Zwane:2017xbg}. Their motivation and even the technicalities are very different with ours.}.

In practice for each value of $\Lambda$ we have two Friedmann equations for normalized Hubble parameter
\begin{eqnarray} \label{H(z)-two}
\begin{cases}
E_1^2(a)=\frac{\Omega_m^{(1)}}{a^3}+\Omega_{\Lambda_1} \\\\
E_2^2(a)=\frac{\Omega_m^{(2)}}{a^3}+\Omega_{\Lambda_2},
\end{cases}
\end{eqnarray}
where we have defined $E_i(a)\equiv H_i(a)/H_0$ where $H_0$ is the Hubble parameter at $a=1$. Now without loss of generality we assume at $a=1$ we will have $E_1(a)$ at work. So we can assume $\Omega_{\Lambda_1}=1-\Omega_m^{(1)}$ for a flat universe. Note that we have effectively $\Lambda$CDM for both regions\footnote{It means in this work we assumed there is no effect by quick jump between $\Omega_{\Lambda_1}$ and $\Omega_{\Lambda_2}$ which of course is an approximation. In future works we will consider the effects of this sharp transition.} which means the perturbation theory is same as standard one (e.g. in \texttt{CosmoMC}). On the other hand since in this work we focus on micro structure in dark energy so we do not expect any changing in matter behavior and we assume $\Omega_m^{(1)}=\Omega_m^{(2)}=\Omega_m$. Hence we remain with two more free parameters, $\Omega_{\Lambda_2}$ and $a_t$, in addition to standard $\Lambda$CDM model. In the following we constrain our free parameters by several sets of data. We also find the best fit of $\Lambda$CDM model with the same data points to make a fair comparison between the two models.

\begin{table}
	\begin{tabular}{|c|c|c|c|c|}
		\hline
		&$\quad$Parameter$\quad$&$\quad$Prior$\quad$&$\quad$Posterior (TT+BAO)$\quad$&$\quad$Posterior (TT+BAO+R19)$\quad$\\
		\hline&&&&\\
		&$\Omega_bh^2$&[0.005\ ,\ 0.1]&$0.02239\pm 0.00023$&$0.02239\pm 0.00022$\\&&&&\\
		
		&$\Omega_ch^2$&[0.001\ ,\ 0.99]&$0.1186\pm 0.0016$&$0.1186\pm 0.0016$\\&&&&\\
		
		\multirow{8}{*}{\rotatebox[origin=c]{90}{$\metalambda$CDM }}&100$\Theta_{MC}$&[0.5\ ,\ 10]&$1.04103\pm 0.00045$&$1.04104\pm 0.00043$\\&&&&\\
		
		&$\tau$&[0.01\ ,\ 0.8]&$0.105^{+0.031}_{-0.027}$&$0.104^{+0.030}_{-0.027}$\\&&&&\\
		
		&$n_s$&[0.8\ ,\ 1.2]&$0.9690\pm 0.0057$&$0.9690\pm 0.0057$\\&&&&\\
		
		&$\ln[10^{10}A_s]$&[2\ ,\ 4]&$3.141^{+0.060}_{-0.052}$&$ 3.140^{+0.059}_{-0.052}$\\&&&&\\
		
		&$a_t$&[0.65\ ,\ 1]&$0.916^{+0.054}_{-0.010}$&$0.916^{+0.055}_{-0.0076}$\\&&&&\\
		
		&$\Omega_{\Lambda_2}$&[0\ ,\ 2]&$0.459^{+0.062}_{-0.082}$&$0.458^{+0.061}_{-0.082}$\\&&&&\\
		
		&$\Omega_{\Lambda_1}$&-&$0.732^{+0.013}_{-0.011}$&$0.732^{+0.013}_{-0.011}$\\&&&&\\
		
		&$H_0\ [\rm km/s/Mpc]$&-&$72.7\pm 1.6$&$72.8\pm 1.6$\\&&&&\\
		
		&&&$\quad\chi^2_{min}=790\quad\rm AIC=806\quad$&$\quad\chi^2_{min}=814\quad\rm AIC=830\quad$\\

		&&&&\\\hline\hline
		&&&&\\
		
		&$\Omega_bh^2$&[0.005\ ,\ 0.1]&$0.02259\pm 0.00021$&$0.02259\pm 0.00021$\\&&&&\\
		
		&$\Omega_ch^2$&[0.001\ ,\ 0.99]&$0.1164\pm 0.0012$&$0.1164\pm 0.0012$\\&&&&\\
		
		\multirow{8}{*}{\rotatebox[origin=c]{90}{$\Lambda$CDM }}&100$\Theta_{MC}$&[0.5\ ,\ 10]&$1.04138\pm 0.00042$&$1.04138\pm 0.00042$\\&&&&\\
		
		&$\tau$&[0.01\ ,\ 0.8]&$ 0.126^{+0.028}_{-0.024}$&$0.126^{+0.028}_{-0.025}$\\&&&&\\
		
		&$n_s$&[0.8\ ,\ 1.2]&$0.9758\pm 0.0049$&$0.9758\pm 0.0049$\\&&&&\\
		
		&$\ln[10^{10}A_s]$&[2\ ,\ 4]&$3.178^{+0.054}_{-0.047}$&$3.177^{+0.054}_{-0.048}$\\&&&&\\
		
		&$\Omega_{\Lambda}$&-&$0.7066\pm0.0069$&$0.7066\pm0.0069$\\&&&&\\
		
		&$H_0\ [\rm km/s/Mpc]$&-&$68.99\pm 0.56$&$68.99\pm 0.56$\\&&&&\\&&&$\quad\chi^2_{min}=795\quad\rm AIC=807\quad$&$\quad\chi^2_{min}=825\quad\rm AIC=837\quad$\\&&&&\\
		
		\hline

	\end{tabular}
	\caption{\label{tab:best-fit}The best fit values for $\Lambda$CDM and $\metalambda$CDM for two sets of data. In addition to $\chi^2$, we have also introduced another measure, Akaike Information Criterion (AIC) as $\text{AIC}=\chi^2_{min}+2N_{model}$, where $N_{model}$ is number of free parameters in the model. The AIC estimator is much better than $\chi^2$ when two models has different numbers of free parameters. A model is more favored among others, if it's AIC has a less value for the same set of the data points. Obviously, in both cases $\metalambda$CDM is more favored than $\Lambda$CDM by using both $\chi^2$ and AIC measures. Obviously, our model is much more consistent with data sets when R19 is included.} \label{back-chi2}
\end{table}

\section{Confronting the $\metalambda$CDM model with observational data sets}\label{datasets}
The data sets with which we have constrained the free parameters of our model are as follows:

\begin{itemize}
	\item Planck 2015 temperature-only $C_\ell^{\rm TT}$ likelihoods for both low$\ell$ and high$\ell$ \cite{Aghanim:2015xee}.
	
	\item BAO volume distance measurements, specifically at $z=0.32$ (LOWZ) \cite{Anderson:2013zyy}, $z=0.57$ (CMASS) \cite{Anderson:2013zyy}, $z=0.106$ (6dFGS) \cite{6df} and $z=0.15$ (MGS) \cite{Ross:2014qpa}.
	
	\item BAO angular diameter distance measurements at $z=0.44$, $z=0.60$ and $z=0.73$ (WiggleZ) \cite{wigglez}.
	
	\item The constraints on the matter power spectrum (mPk) at $z=0.35$ from  SDSS DR4 luminous red galaxies (LRG) \cite{Tegmark:2006az}. 
	
	\item The latest measurement of $H_0$ by Riess \textit{et al.} \cite{Riess:2019cxk}. We refer to this data point as R19.
\end{itemize}

We have implemented our modifications into the publicly available code \texttt{CAMB} \cite{camb,Lewis:1999bs}, in order to calculate the theoretical predictions of $\metalambda$CDM model for the observables described above. Note that we've just modified the background evolution of the universe according to our model and we did not change the equations of \texttt{CAMB} at the level of perturbation. This is because we have not added a field to the contents of cosmos; what we have added is another cosmological constant that does not cluster by definition. 

\begin{figure}
	\centering
	\includegraphics[width=1\linewidth]{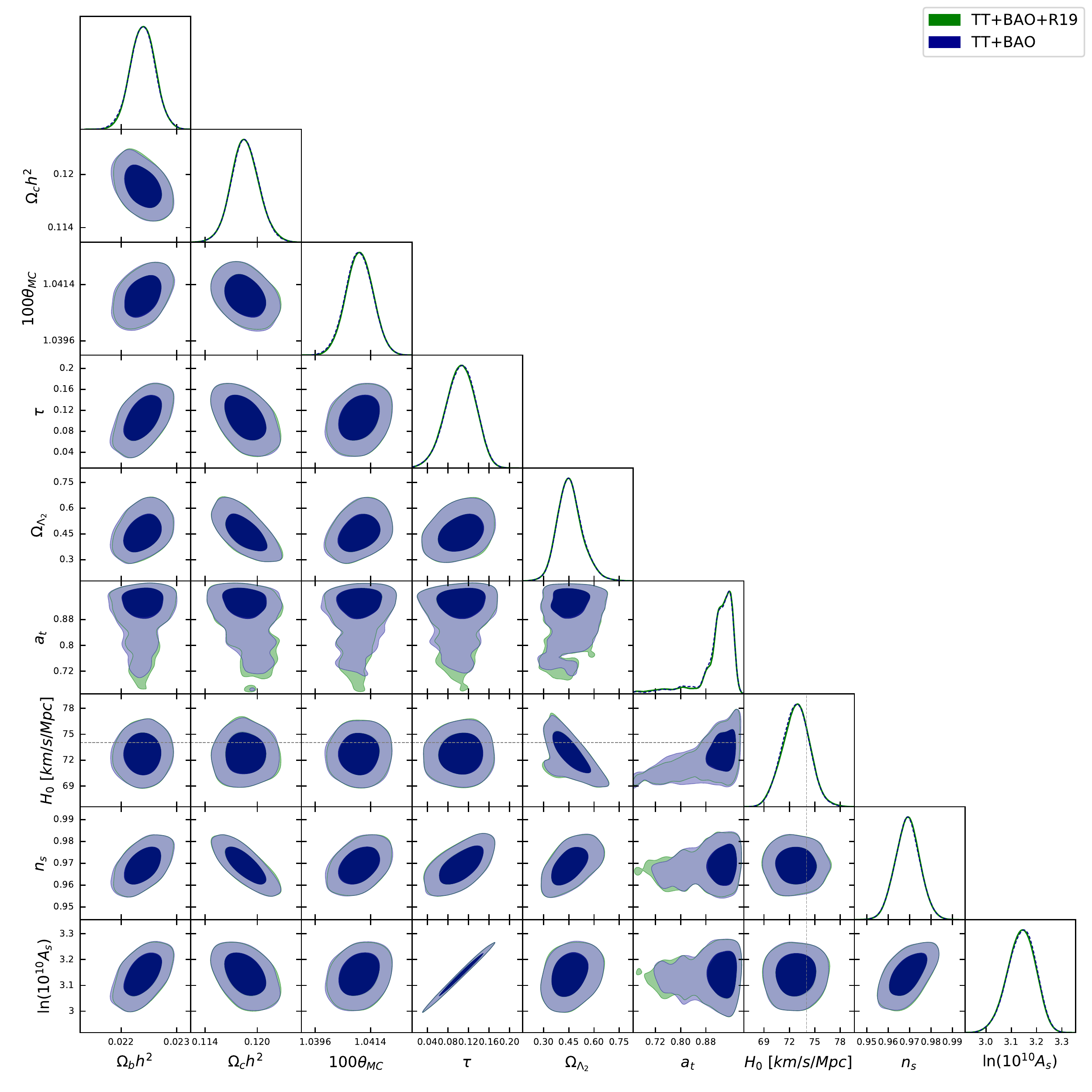}
	\caption{$\metalambda$CDM (with non-trivial $a_t$) $68\%$ and $95\%$ parameter constraint contours from two sets of data points both including CMB and BAO but the green one has R19 in addition. We have also added a marker in $H_0$ posterior distribution and corresponding contours to show the value of $H_0$ data point. we have to emphasize that $\metalambda$CDM model prediction for $H_0$  for only CMB+BAO dataset is consistent with R19. This is very important to reach to this consistency before adding R19 as a data point since for sure adding R19 is in favor of higher $H_0$.}
	\label{fig:9triangle}
\end{figure}

\begin{figure}
	\centering
	\includegraphics[width=0.5\linewidth]{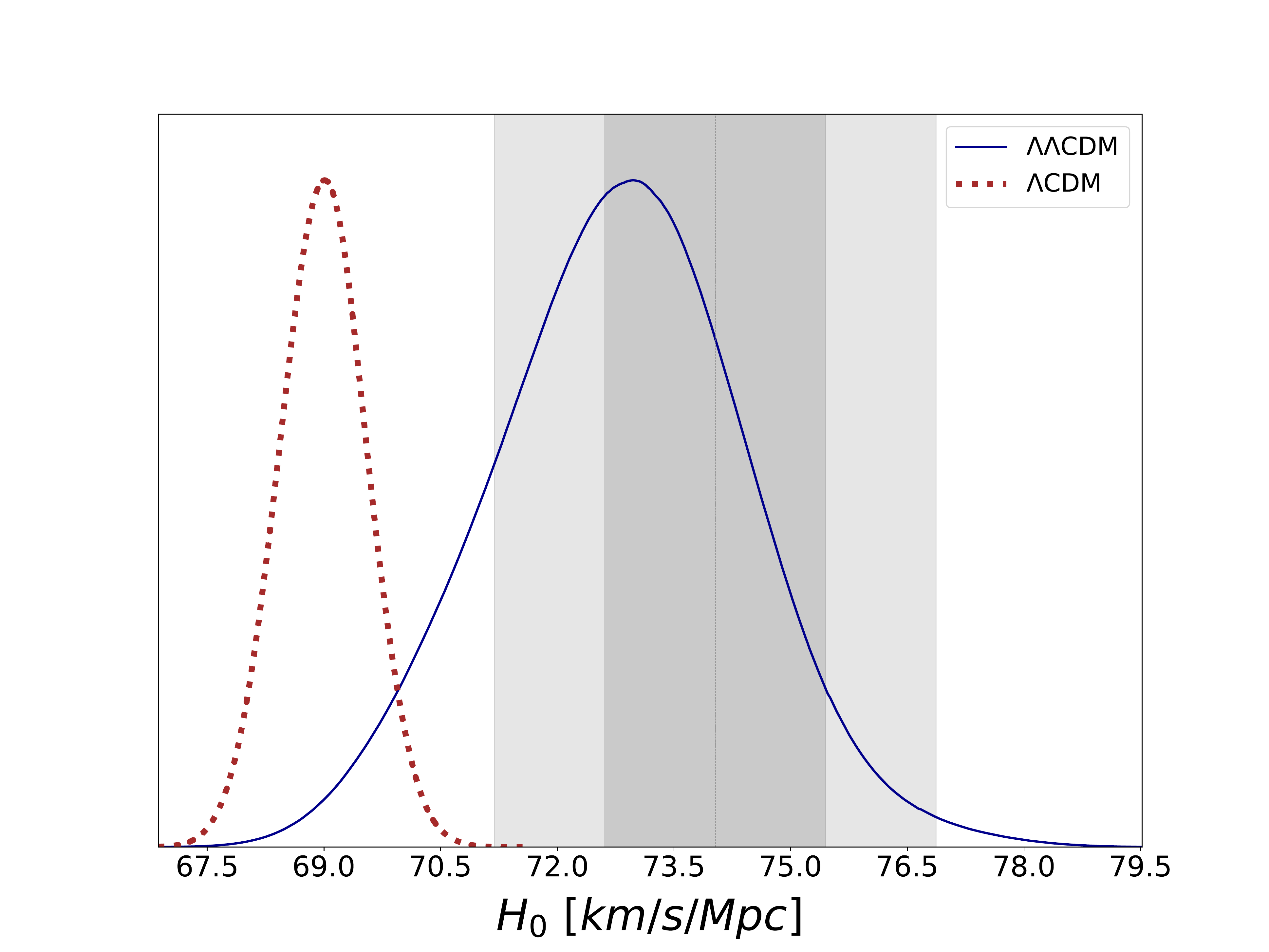}
	\caption{Comparison between the inferred $H_0$ posteriors from two models, $\Lambda$CDM and $\metalambda$CDM when all the data points (TT+BAO+R19) are used and that of directly measured by R19. It is clear that our model solves the $H_0$ tension since $\metalambda$CDM posterior for $H_0$ parameter has overlap with the same posterior from $\Lambda$CDM model in $2\sigma$ region. In addition R19 is obviously compatible with $\metalambda$CDM model prediction.}
	\label{fig:h0}
\end{figure}

Using the famous Monte Carlo Markov Chain code, \texttt{CosmoMC} \cite{Lewis:2002ah,Lewis:1999bs}, we have sampled the parameter space of this model and put constraints on them. The parameter space that we have constrained is

\begin{equation}
\mathcal{P}=\{\Omega_bh^2,\Omega_ch^2,100\Theta_{MC},\tau,n_s,\ln[10^{10}A_s],a_t,\Omega_{\Lambda_2}\},
\end{equation} 
where $\Omega_bh^2$ is the today physical baryon density; $\Omega_ch^2$ is the today physical cold dark matter density; $\Theta$ is the ratio of sound horizon at the time of decoupling to the angular diameter distance of last scattering surface; $\tau$ is the reionization optical depth; $n_s$ is the scalar spectral index; $A_s$ denotes the amplitude of primordial scalar power spectrum; $a_t$ is the scale factor at which transition happens and $\Omega_{\Lambda_2}$ is the second value of dark energy density before transition. We have set flat priors for all of these parameters. In order to have a fair model comparison, we have also investigated the statistical situation of the flat $\Lambda$CDM with the same priors and using the same data sets and methods. A summary of the results can be found in TABLE \ref{tab:best-fit}. In addition parameter constraint contours and posterior distributions for $\metalambda$CDM are shown in the FIG. \ref{fig:9triangle}. In  order to compare the ability of two models in dealing with the directly measured $H_0$, we have plotted FIG. \ref{fig:h0}. We have to mention that we have put a prior on $a_t$ as $[0.65,1]$ and its reason is discussed in the Appendix \ref{at}.

\begin{figure}
\begingroup
\makeatletter
\providecommand\color[2][]{%
	\GenericError{(gnuplot) \space\space\space\@spaces}{%
		Package color not loaded in conjunction with
		terminal option `colourtext'%
	}{See the gnuplot documentation for explanation.%
	}{Either use 'blacktext' in gnuplot or load the package
		color.sty in LaTeX.}%
	\renewcommand\color[2][]{}%
}%
\providecommand\includegraphics[2][]{%
	\GenericError{(gnuplot) \space\space\space\@spaces}{%
		Package graphicx or graphics not loaded%
	}{See the gnuplot documentation for explanation.%
	}{The gnuplot epslatex terminal needs graphicx.sty or graphics.sty.}%
	\renewcommand\includegraphics[2][]{}%
}%
\providecommand\rotatebox[2]{#2}%
\@ifundefined{ifGPcolor}{%
	\newif\ifGPcolor
	\GPcolortrue
}{}%
\@ifundefined{ifGPblacktext}{%
	\newif\ifGPblacktext
	\GPblacktexttrue
}{}%
\let\gplgaddtomacro\g@addto@macro
\gdef\gplbacktext{}%
\gdef\gplfronttext{}%
\makeatother
\ifGPblacktext
\def\colorrgb#1{}%
\def\colorgray#1{}%
\else
\ifGPcolor
\def\colorrgb#1{\color[rgb]{#1}}%
\def\colorgray#1{\color[gray]{#1}}%
\expandafter\def\csname LTw\endcsname{\color{white}}%
\expandafter\def\csname LTb\endcsname{\color{black}}%
\expandafter\def\csname LTa\endcsname{\color{black}}%
\expandafter\def\csname LT0\endcsname{\color[rgb]{1,0,0}}%
\expandafter\def\csname LT1\endcsname{\color[rgb]{0,1,0}}%
\expandafter\def\csname LT2\endcsname{\color[rgb]{0,0,1}}%
\expandafter\def\csname LT3\endcsname{\color[rgb]{1,0,1}}%
\expandafter\def\csname LT4\endcsname{\color[rgb]{0,1,1}}%
\expandafter\def\csname LT5\endcsname{\color[rgb]{1,1,0}}%
\expandafter\def\csname LT6\endcsname{\color[rgb]{0,0,0}}%
\expandafter\def\csname LT7\endcsname{\color[rgb]{1,0.3,0}}%
\expandafter\def\csname LT8\endcsname{\color[rgb]{0.5,0.5,0.5}}%
\else
\def\colorrgb#1{\color{black}}%
\def\colorgray#1{\color[gray]{#1}}%
\expandafter\def\csname LTw\endcsname{\color{white}}%
\expandafter\def\csname LTb\endcsname{\color{black}}%
\expandafter\def\csname LTa\endcsname{\color{black}}%
\expandafter\def\csname LT0\endcsname{\color{black}}%
\expandafter\def\csname LT1\endcsname{\color{black}}%
\expandafter\def\csname LT2\endcsname{\color{black}}%
\expandafter\def\csname LT3\endcsname{\color{black}}%
\expandafter\def\csname LT4\endcsname{\color{black}}%
\expandafter\def\csname LT5\endcsname{\color{black}}%
\expandafter\def\csname LT6\endcsname{\color{black}}%
\expandafter\def\csname LT7\endcsname{\color{black}}%
\expandafter\def\csname LT8\endcsname{\color{black}}%
\fi
\fi
\setlength{\unitlength}{0.0500bp}%
\ifx\gptboxheight\undefined%
\newlength{\gptboxheight}%
\newlength{\gptboxwidth}%
\newsavebox{\gptboxtext}%
\fi%
\setlength{\fboxrule}{0.5pt}%
\setlength{\fboxsep}{1pt}%
\begin{picture}(5100.00,3400.00)%
\gplgaddtomacro\gplbacktext{%
	\csname LTb\endcsname
	\put(747,595){\makebox(0,0)[r]{\strut{}$0.85$}}%
	\csname LTb\endcsname
	\put(747,1031){\makebox(0,0)[r]{\strut{}$0.9$}}%
	\csname LTb\endcsname
	\put(747,1468){\makebox(0,0)[r]{\strut{}$0.95$}}%
	\csname LTb\endcsname
	\put(747,1904){\makebox(0,0)[r]{\strut{}$1$}}%
	\csname LTb\endcsname
	\put(747,2340){\makebox(0,0)[r]{\strut{}$1.05$}}%
	\csname LTb\endcsname
	\put(747,2777){\makebox(0,0)[r]{\strut{}$1.1$}}%
	\csname LTb\endcsname
	\put(747,3213){\makebox(0,0)[r]{\strut{}$1.15$}}%
	\csname LTb\endcsname
	\put(849,409){\makebox(0,0){\strut{}$0$}}%
	\csname LTb\endcsname
	\put(1638,409){\makebox(0,0){\strut{}$0.5$}}%
	\csname LTb\endcsname
	\put(2427,409){\makebox(0,0){\strut{}$1$}}%
	\csname LTb\endcsname
	\put(3215,409){\makebox(0,0){\strut{}$1.5$}}%
	\csname LTb\endcsname
	\put(4004,409){\makebox(0,0){\strut{}$2$}}%
	\csname LTb\endcsname
	\put(4793,409){\makebox(0,0){\strut{}$2.5$}}%
}%
\gplgaddtomacro\gplfronttext{%
	\csname LTb\endcsname
	\put(153,1904){\rotatebox{-270}{\makebox(0,0){\strut{}$(D_M/r_d)/(D_M/r_d)_{\Lambda\rm CDM}$}}}%
	\csname LTb\endcsname
	\put(2821,130){\makebox(0,0){\strut{}$z$}}%
	\csname LTb\endcsname
	\put(4005,3046){\makebox(0,0)[r]{\strut{}\scriptsize{$\metalambda$CDM (TT+BAO+R19)}}}%
	\csname LTb\endcsname
	\put(4005,2860){\makebox(0,0)[r]{\strut{}\scriptsize{$\metalambda$CDM (TT+BAO)}}}%
	\csname LTb\endcsname
	\put(3600,1660){\makebox(0,0)[r]{\strut{}\tiny{SDSS quasars}}}%
	\csname LTb\endcsname
	\put(4500,1060){\makebox(0,0)[r]{\strut{}\tiny{Ly-$\alpha$}}}%
	\csname LTb\endcsname
	\put(2250,1460){\makebox(0,0)[r]{\strut{}\tiny{DES}}}%
	\csname LTb\endcsname
	\put(1750,2200){\makebox(0,0)[r]{\strut{}\tiny{BOSS DR12}}}%

}%
\gplbacktext
\put(0,0){\includegraphics{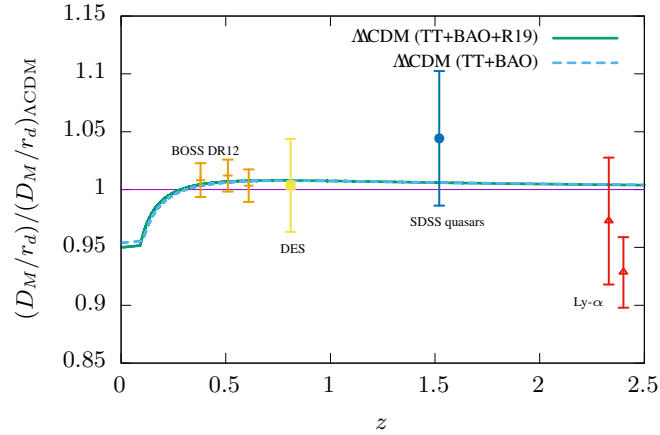}}%
\gplfronttext
\end{picture}%
\endgroup

	\caption{The metric distance, $ D_M(z)$, normalized to $\Lambda$CDM best fit values prediction.}
	\label{fig:D-M}
\end{figure}

\begin{figure}
	\begingroup
	\makeatletter
	\providecommand\color[2][]{%
		\GenericError{(gnuplot) \space\space\space\@spaces}{%
			Package color not loaded in conjunction with
			terminal option `colourtext'%
		}{See the gnuplot documentation for explanation.%
		}{Either use 'blacktext' in gnuplot or load the package
			color.sty in LaTeX.}%
		\renewcommand\color[2][]{}%
	}%
	\providecommand\includegraphics[2][]{%
		\GenericError{(gnuplot) \space\space\space\@spaces}{%
			Package graphicx or graphics not loaded%
		}{See the gnuplot documentation for explanation.%
		}{The gnuplot epslatex terminal needs graphicx.sty or graphics.sty.}%
		\renewcommand\includegraphics[2][]{}%
	}%
	\providecommand\rotatebox[2]{#2}%
	\@ifundefined{ifGPcolor}{%
		\newif\ifGPcolor
		\GPcolortrue
	}{}%
	\@ifundefined{ifGPblacktext}{%
		\newif\ifGPblacktext
		\GPblacktexttrue
	}{}%
	\let\gplgaddtomacro\g@addto@macro
	\gdef\gplbacktext{}%
	\gdef\gplfronttext{}%
	\makeatother
	\ifGPblacktext
	\def\colorrgb#1{}%
	\def\colorgray#1{}%
	\else
	\ifGPcolor
	\def\colorrgb#1{\color[rgb]{#1}}%
	\def\colorgray#1{\color[gray]{#1}}%
	\expandafter\def\csname LTw\endcsname{\color{white}}%
	\expandafter\def\csname LTb\endcsname{\color{black}}%
	\expandafter\def\csname LTa\endcsname{\color{black}}%
	\expandafter\def\csname LT0\endcsname{\color[rgb]{1,0,0}}%
	\expandafter\def\csname LT1\endcsname{\color[rgb]{0,1,0}}%
	\expandafter\def\csname LT2\endcsname{\color[rgb]{0,0,1}}%
	\expandafter\def\csname LT3\endcsname{\color[rgb]{1,0,1}}%
	\expandafter\def\csname LT4\endcsname{\color[rgb]{0,1,1}}%
	\expandafter\def\csname LT5\endcsname{\color[rgb]{1,1,0}}%
	\expandafter\def\csname LT6\endcsname{\color[rgb]{0,0,0}}%
	\expandafter\def\csname LT7\endcsname{\color[rgb]{1,0.3,0}}%
	\expandafter\def\csname LT8\endcsname{\color[rgb]{0.5,0.5,0.5}}%
	\else
	\def\colorrgb#1{\color{black}}%
	\def\colorgray#1{\color[gray]{#1}}%
	\expandafter\def\csname LTw\endcsname{\color{white}}%
	\expandafter\def\csname LTb\endcsname{\color{black}}%
	\expandafter\def\csname LTa\endcsname{\color{black}}%
	\expandafter\def\csname LT0\endcsname{\color{black}}%
	\expandafter\def\csname LT1\endcsname{\color{black}}%
	\expandafter\def\csname LT2\endcsname{\color{black}}%
	\expandafter\def\csname LT3\endcsname{\color{black}}%
	\expandafter\def\csname LT4\endcsname{\color{black}}%
	\expandafter\def\csname LT5\endcsname{\color{black}}%
	\expandafter\def\csname LT6\endcsname{\color{black}}%
	\expandafter\def\csname LT7\endcsname{\color{black}}%
	\expandafter\def\csname LT8\endcsname{\color{black}}%
	\fi
	\fi
	\setlength{\unitlength}{0.0500bp}%
	\ifx\gptboxheight\undefined%
	\newlength{\gptboxheight}%
	\newlength{\gptboxwidth}%
	\newsavebox{\gptboxtext}%
	\fi%
	\setlength{\fboxrule}{0.5pt}%
	\setlength{\fboxsep}{1pt}%
	\begin{picture}(5100.00,3400.00)%
	\gplgaddtomacro\gplbacktext{%
		\csname LTb\endcsname
		\put(747,595){\makebox(0,0)[r]{\strut{}$0.85$}}%
		\csname LTb\endcsname
		\put(747,1031){\makebox(0,0)[r]{\strut{}$0.9$}}%
		\csname LTb\endcsname
		\put(747,1468){\makebox(0,0)[r]{\strut{}$0.95$}}%
		\csname LTb\endcsname
		\put(747,1904){\makebox(0,0)[r]{\strut{}$1$}}%
		\csname LTb\endcsname
		\put(747,2340){\makebox(0,0)[r]{\strut{}$1.05$}}%
		\csname LTb\endcsname
		\put(747,2777){\makebox(0,0)[r]{\strut{}$1.1$}}%
		\csname LTb\endcsname
		\put(747,3213){\makebox(0,0)[r]{\strut{}$1.15$}}%
		\csname LTb\endcsname
		\put(849,409){\makebox(0,0){\strut{}$0$}}%
		\csname LTb\endcsname
		\put(1638,409){\makebox(0,0){\strut{}$0.2$}}%
		\csname LTb\endcsname
		\put(2427,409){\makebox(0,0){\strut{}$0.4$}}%
		\csname LTb\endcsname
		\put(3215,409){\makebox(0,0){\strut{}$0.6$}}%
		\csname LTb\endcsname
		\put(4004,409){\makebox(0,0){\strut{}$0.8$}}%
		\csname LTb\endcsname
		\put(4793,409){\makebox(0,0){\strut{}$1$}}%
	}%
	\gplgaddtomacro\gplfronttext{%
		\csname LTb\endcsname
		\put(153,1904){\rotatebox{-270}{\makebox(0,0){\strut{}$(D_V/r_d)/(D_V/r_d)_{\Lambda \rm CDM}$}}}%
		\csname LTb\endcsname
		\put(2821,130){\makebox(0,0){\strut{}$z$}}%
		\csname LTb\endcsname
		\put(4005,3046){\makebox(0,0)[r]{\strut{}\scriptsize{$\metalambda$CDM (TT+BAO+R19)}}}%
		\csname LTb\endcsname
		\put(4005,2860){\makebox(0,0)[r]{\strut{}\scriptsize{$\metalambda$CDM (TT+BAO)}}}%
		\csname LTb\endcsname
		\put(3650,2400){\makebox(0,0)[r]{\strut{}\tiny{WiggleZ}}}%
		\csname LTb\endcsname
		\put(4000,1300){\makebox(0,0)[r]{\strut{}\tiny{SDSS LRG}}}%
		\csname LTb\endcsname
		\put(1700,2900){\makebox(0,0)[r]{\strut{}\tiny{SDSS MGS}}}%
		\csname LTb\endcsname
		\put(1450,1300){\makebox(0,0)[r]{\strut{}\tiny{6dFGS}}}%
		
	}%
	\gplbacktext
	\put(0,0){\includegraphics{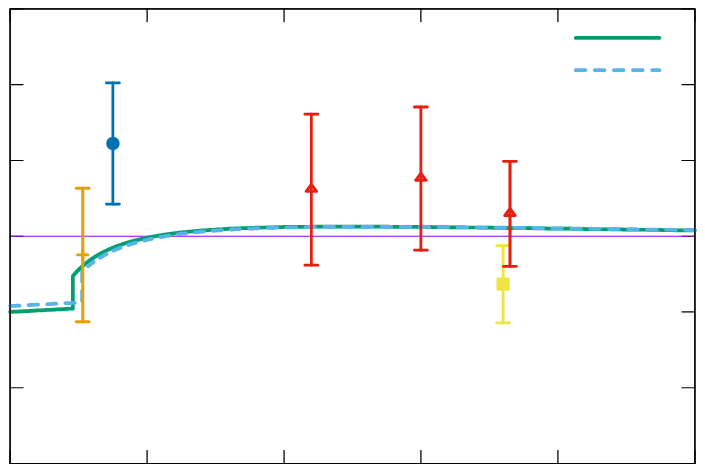}}%
	\gplfronttext
	\end{picture}%
	\endgroup
	
	\caption{Here we plot the volume distance, $D_V$,  normalized to $\Lambda$CDM best fit values
		 prediction.}\label{fig:D-V}
\end{figure}

We have also plotted metric distance and volume distance versus redshift for our best fits in FIG. \ref{fig:D-M} and FIG. \ref{fig:D-V} respectively to have a sense about the behavior of our model and compare with that of $\Lambda$CDM.

\section{Concluding remarks}\label{conclusions}
Based on the structure of $H_0$ tension, we proposed a dark energy model which says dark energy underwent a phase transition in its history. In this work, our idea has been realized by the simplest scenario: instead of a cosmological constant we have two distinctive  values for the cosmological constant and we named our model:  $\metalambda$CDM. In addition, inspired by the Ising model, we supposed we have two different behaviors before and after a critical temperature (which corresponds to a transition scale factor in cosmology). Before the transition, the universe switches between $\Lambda_1$ and $\Lambda_2$ (with fractional densities of 0.73 and 0.46 respectively) while it settles into the standard $\Lambda$CDM model after the transition scale factor, $a_t\approx0.916$. We'd like to mention that our best fit for the transition scale factor is pretty consistent with the transition redshift reported in \cite{Benevento:2020fev}.

We have checked our model by considering the both high$\ell$ and low$\ell$ CMB Temperature power spectrum, BAO's, matter power spectrum at $z=0.35$ from luminous red galaxies and the most recent $H_0$ measurement. We summarized the results in TABLE \ref{tab:best-fit} which shows less $\chi^2$ for our model: $\Delta\chi^2=-11$ and $\Delta\chi^2=-5$ with and without $H_0$ data point. However $\chi^2$ analysis is not a good estimator when two models have different number of free parameters. Instead we have used AIC which penalizes the model with more free parameters. In this estimator our $\metalambda$CDM model is better than $\Lambda$CDM by having relative AIC as $\Delta AIC=-1$ and $\Delta AIC=-7$, without and with R19 in datasets respectively. This means AIC estimator prefers our model even if it has more free parameters.

\section{Future Perspectives}\label{future}

We think the idea of phase transition in dark sector is a very rich concept both phenomenologically and theoretically. This idea is supported with the way that we understand the cosmological tensions: all of these tensions can be phrased as inconsistencies between early and late time physics and so a phase transitions in mid redshifts can address the different  behaviors of the universe in early and late times. A phase-transition in dark energy has a very interesting deep consequence: dark energy has (micro-)structures.

This idea can be checked phenomenologically by checking the bare observations and see if there is a kind of different behaviors for cosmological parameters in different redshifts. For example as we mentioned above the behavior of $H(z)$ is different for low and high redshift as it is reported in \cite{Bernal:2016gxb}. In addition in \cite{Zhao:2017cud} the behavior of equation of state of dark energy seems is not $w=-1$ and it oscillates. This is also in agreement with our idea where we assume dark energy switches between two different values. However the frequency of oscillations is very larger in our model and we should check our model for lower frequencies too in future works.

In the theoretical side is a vast era of exploration: in this work we focused on the simplest scenario inspired by the Ising model. We will generalize our approach for more precise models e.g. by removing fast phase transition. In addition we can think about other models e.g. Heisenberg model, Potts model and etc. One way to think about this idea is working in a continuum regime which is remained for the future work.

\vspace{.5 cm}
\textit{Acknowledgments:}
We are grateful to S. Baghram, M. Farhang and S.M.S. Movahed for fruitful discussions as well as their comments on the early draft. We also thank A. Hosseiny and  B. Mostaghel for useful discussions. We thank H. Moshafi for his guidance about  \texttt{CosmoMC}. This work is supported by Iran National Science Foundation (INSF), project no. 98022568. NK thanks School of Physics at IPM where he is a part time researcher.

\appendix
\section{A comment on $a_t$}\label{at} 
\begin{figure}[h]
	\centering
	\includegraphics[width=0.6\linewidth]{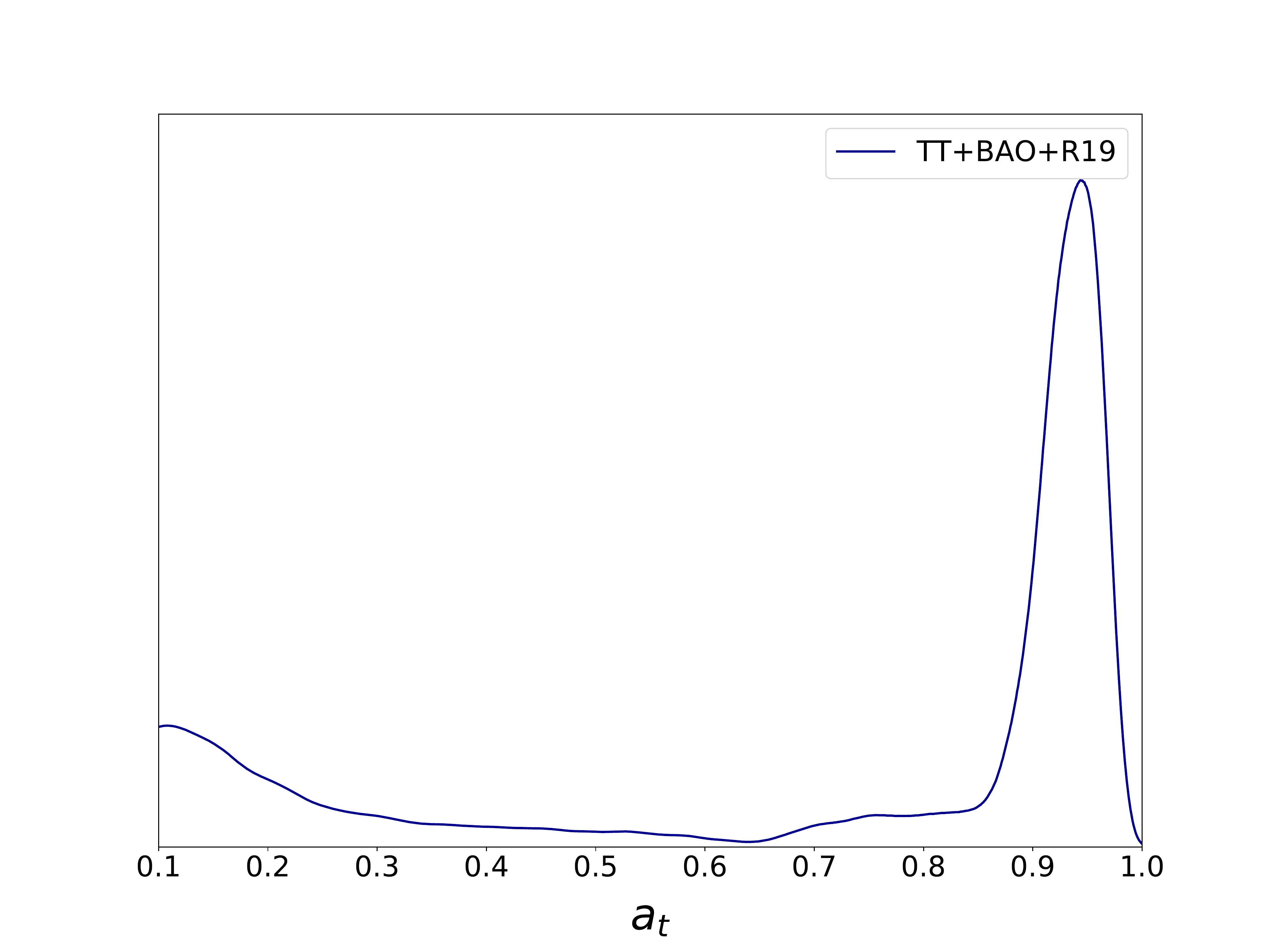}
	\caption{Posterior probability distribution for $a_t$. It seems there is another chance for transition to occur at high redshifts, $z\approx10$.}
	\label{fig:atr}
\end{figure}

\begin{figure}
	\centering
	\includegraphics[width=0.6\linewidth]{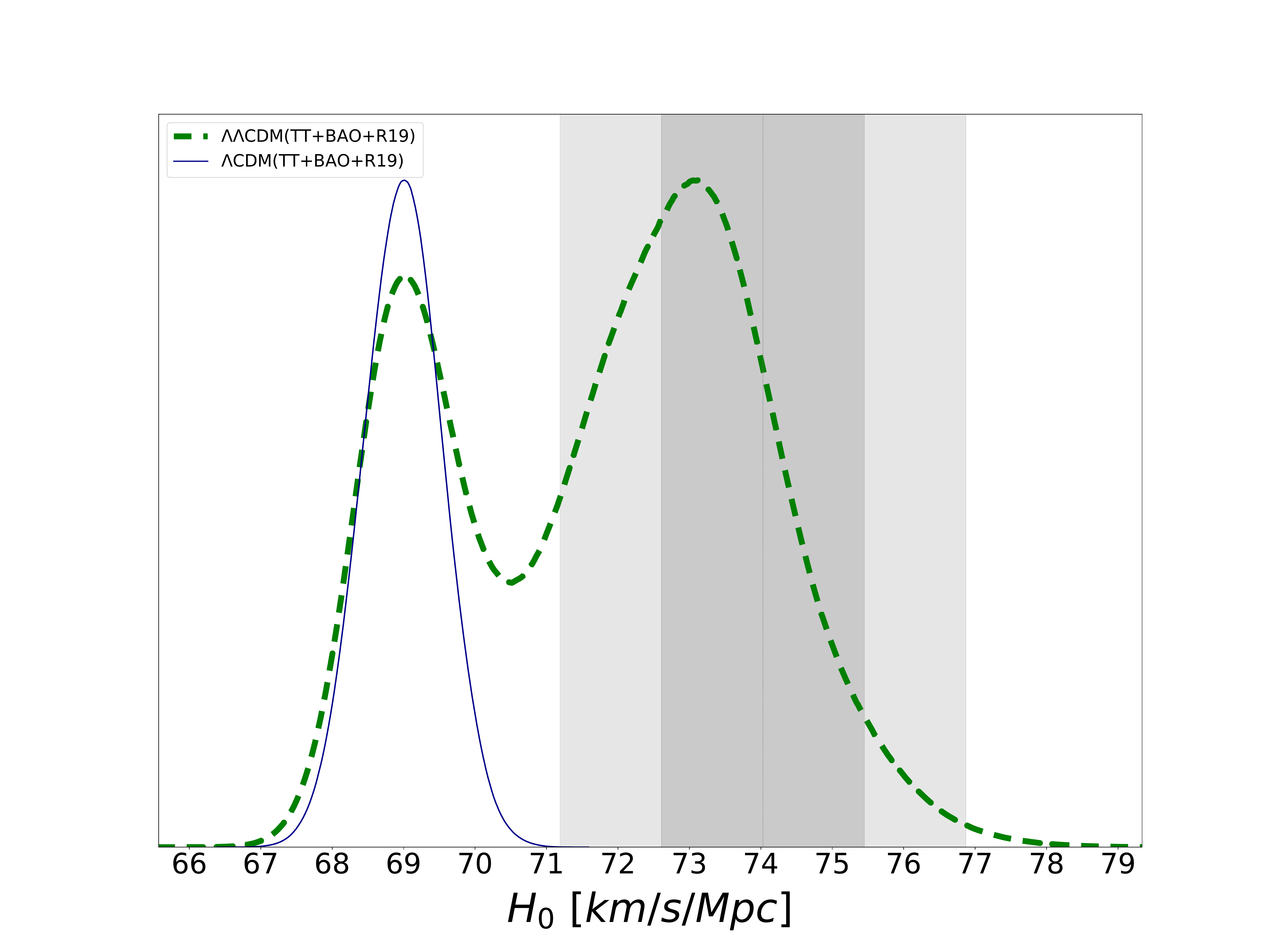}
	\caption{Posterior probability distribution for $H_0$ where there is no limit on prior of $a_t$. It is evident that the value for the second peak ($a_t\sim 0.1$) of $\metalambda$CDM coincides with the best fit value of $\Lambda$CDM, see FIG. \ref{fig:h0}. The gray bands show the value of directly measured $H_0$ \cite{Riess:2019cxk} up to two standard deviations.}
	\label{fig:ch0}
\end{figure}

When we were trying to constrain the free parameters of $\metalambda$CDM, first we didn't put bounds on $a_t$ prior and noticed that there are two peaks in the 1-d marginalized posterior for this parameter, cf. FIG. \ref{fig:atr}.
The $\chi^2$ corresponding to the second peak ($a_t\sim 0.1$) was more or less as much as the $\chi^2_{min}$ for $\Lambda$CDM which is understandable: our model is designed to become $\Lambda$CDM after transition and since data is insensitive to the early time behavior of dark energy\footnote{Among the data we have used, CMB temperature power spectrum is sensitive to dark energy via late ISW effect and all the other data points belong to late time i.e $z\le1$.}, our model is not distinguishable from $\Lambda$CDM when the transition occurs at high redshifts. So we decided to put a limit on the prior of $a_t$ to be $[0.65,1]$, because we thought only the higher peak in FIG. \ref{fig:atr} leads to new physics. Otherwise, we would have a two-peak posterior for other parameters. For instance in FIG. \ref{fig:ch0} the situation for $H_0$ is depicted.

\end{document}